\PassOptionsToPackage{table}{xcolor}
\documentclass[preprint,12pt,review]{elsarticle}

\usepackage[numbers]{natbib}

\usepackage{float}
\usepackage{graphicx}
\usepackage{tabularx}
\usepackage{amssymb}
\usepackage{color}
\usepackage{subfigure}
\usepackage{array}
\usepackage{lscape}
\usepackage{dcolumn,longtable,hhline,colortbl}
\usepackage[rotateright]{rotating}
\usepackage{booktabs,makecell,multirow}
\usepackage[section]{placeins}
\usepackage{enumerate}
\usepackage{color,colortbl}

\usepackage{lineno}
\usepackage{color}
\usepackage{soul}
\usepackage{url}
\usepackage[table]{xcolor}
\usepackage{makecell}
\usepackage{amsthm}
\usepackage{multirow}
\usepackage{bm}
\usepackage{autoaligne}
\usepackage{mathtools}
\usepackage[flushleft]{threeparttable}
\usepackage{setspace}

\biboptions{sort&compress}

\usepackage{nomencl}
\makenomenclature
\usepackage[linesnumbered,ruled,vlined,algo2e]{algorithm2e}

\usepackage{tabularray}
\usepackage{multirow}
\usepackage{algpseudocode}
\usepackage{amssymb}
\usepackage{amsfonts}

\usepackage{algorithm,algcompatible,amsmath}
\usepackage{hyperref}

\usepackage[defaultcolor=red]{changes}  
\definechangesauthor[name={Author}, color=red]{A} 

\algnewcommand\INPUT{\item[\textbf{Input:}]}
\algnewcommand\OUTPUT{\item[\textbf{Output:}]}

\usepackage{amssymb}

\journal{Advanced Engineering Informatics}

\begin{document}
\begin{sloppypar}
\begin{frontmatter}

\title{A Multidisciplinary Design and Optimization (MDO) Agent Driven by Large Language Models}

\address[add1]{State Key Laboratory of Fluid Power Transmission and Control, Zhejiang University, Hangzhou, 310027, PR China}

\address[add3]{School of Aeronautics and Astronautics, Zhejiang University, Hangzhou 310027, PR China} 

\address[add4]{China Ship Development and Design Center, Wuhan, 430064, PR China}

\author[add1]{Bingkun Guo}
\author[add1]{Wentian Li}
\author[add1]{Xiaojian Liu}
\author[add3]{Jiaqi Luo}
\author[add4]{Zibin Yu}
\author[add4]{Dalong Dong}

\author[add1]{Shuyou Zhang}
\author[add1]{Yiming Zhang\corref{cor1}}

\cortext[cor1]{Corresponding author: yimingzhang@zju.edu.cn}

\begin{abstract}
To accelerate mechanical design and enhance design quality and innovation, we present a Multidisciplinary Design and Optimization (MDO) Agent driven by Large Language Models (LLMs). The agent semi‑automates the end‑to‑end workflow by orchestrating three core capabilities: (i) natural‑language–driven parametric modeling, (ii) retrieval‑augmented generation (RAG) for knowledge‑grounded conceptualization, and (iii) intelligent orchestration of engineering software for performance verification and optimization. Working in tandem, these capabilities interpret high‑level, unstructured intent, translate it into structured design representations, automatically construct parametric 3D CAD models, generate reliable concept variants using external knowledge bases, and conduct evaluation with iterative optimization via tool calls such as finite‑element analysis (FEA). Validation on three representative cases—a gas‑turbine blade, a machine‑tool column, and a fractal heat sink—shows that the agent completes the pipeline from natural‑language intent to verified and optimized designs with reduced manual scripting and setup effort, while promoting innovative design exploration. This work points to a practical path toward human–AI collaborative mechanical engineering and lays a foundation for more dependable, vertically customized MDO systems.

\end{abstract}




\end{frontmatter}




\section{Introduction}\label{introduction}
Mechanical structural design remains central to modern engineering, yet its practice is still dominated by expert‑centric, tool‑fragmented workflows \cite{pereira2022review}. The conventional workflow is an iterative and labor-intensive process heavily reliant on engineering expertise and facilitated by computer-aided design (CAD) and computer-aided engineering (CAE) tools \cite{wang2024computer}. However, this human-centric paradigm suffers from significant limitations. It is often constrained by the empirical knowledge of designers, leading to prolonged development cycles and stifling innovation. Moreover, a fragmented toolchain and a persistent "semantic gap" in human-computer interaction hinder efficiency, creating data silos, and failing to fully capture complex design intent \cite{zhao2024explainability}. These limitations call for an interaction paradigm that faithfully captures high‑level intent while translating it into executable, verifiable pipelines.

Recent advances in Large Language Models (LLMs) offer a credible route to soften these bottlenecks \cite{annepaka2025large}. With remarkable capabilities in natural language understanding, reasoning, and code generation, LLMs can serve as an intuitive interface for complex engineering tasks \cite{kartashov2025large}. They have the potential to interpret high-level commands into executable modeling scripts, explore wider design spaces, and ultimately foster innovation \cite{vaithilingam2022expectation}. A significant research gap exists in the systematic orchestration of these capabilities across the entire rigorous engineering workflow. However, turning such capabilities into dependable, end‑to‑end MDO pipelines requires clarifying what LLMs can reliably provide and how they should be orchestrated across design, analysis, and optimization.

We therefore briefly outline the technical underpinnings of LLMs to motivate their suitability for planning, code synthesis, and tool use in engineering. Large Language Models represent a significant breakthrough in artificial intelligence, demonstrating transformative potential across various domains \cite{chang2024survey}. Current mainstream LLMs are primarily based on the Transformer architecture, particularly causal decoder models. Through unsupervised pre-training on massive text datasets, they learn general linguistic patterns, world knowledge, and preliminary reasoning capabilities \cite{mckinzie2024mm1}. Subsequent post-training, involving methods like Supervised Fine-Tuning (SFT) and Reinforcement Learning from Human Feedback (RLHF), further optimizes model performance for specific downstream tasks and aligns their outputs with human preferences \cite{naveed2023comprehensive}. LLMs possess core capabilities encompassing exceptional natural language understanding, contextual reasoning, knowledge integration, and powerful code generation. These abilities have led to remarkable achievements in numerous Natural Language Processing sub-fields and show immense promise in code generation, enabling the translation of natural language requirements into executable code. These properties make LLMs a plausible hub for mapping natural‑language intent to programmatic control—particularly appealing for code‑driven CAD \cite{gao2024large}.

To overcome the inherent limitations of traditional Graphical User Interface (GUI)‑based approaches in terms of automation and reusability, code-driven modeling has emerged as a significant paradigm in the field of CAD. Unlike manual GUI interactions, this paradigm employs programming languages to precisely define the construction logic and parametric relationships of geometric entities, thereby enabling a high degree of automation and parametric control \cite{machado2019parametric}. This method allows designers to rapidly generate model variations by modifying code and inherently supports version control, which greatly facilitates design iteration, collaboration, and knowledge reuse. Yet textual scripts alone do not resolve downstream analysis and optimization complexity, where expert configuration and fine‑grained spatial reasoning remain demanding.

The implementation of code-driven modeling varies, with common approaches including the use of general-purpose scripting languages like Python to access the Application Programming Interfaces (APIs) of CAD systems, or the adoption of specialized parametric modeling languages. For instance, OpenSCAD offers a concise, declarative language for solid modeling, while Python-based libraries such as CadQuery provide more expressive, fluent APIs that make parametric scripts more readable and maintainable. Nevertheless, despite the unparalleled precision and automation potential of code-driven modeling, it typically imposes a steep learning curve, requiring designers to possess solid programming skills. Furthermore, the purely textual representation lacks geometric intuitiveness, making it challenging to quickly map complex design intent to the underlying code logic. Consequently, there is a compelling need to explore a new interaction paradigm that synergizes the intuitiveness of natural language with the precision of code-based control.

Downstream verification and optimization further amplify the challenge. Finite Element Analysis is an indispensable technique for validating the performance of engineering designs, while structural optimization is the subsequent critical step for identifying the best possible solution. Established commercial software packages, such as ANSYS and ABAQUS \cite{cui2024scaled}, offer powerful simulation capabilities to accurately predict a structure's mechanical behavior under complex operating conditions. Concurrently, a wide array of optimization algorithms (e.g., Genetic Algorithms, Particle Swarm Optimization) and integrated optimization tools provide systematic methodologies for enhancing design performance \cite{sheng2025isogeometric}. 

Despite their power, the use of these tools in conventional workflows is often encumbered by a high barrier to entry and significant procedural complexity. A typical FEA process involves tedious manual operations, including geometry cleanup, mesh generation, and the precise application of boundary conditions, demanding substantial expertise from the engineer. Similarly, configuring and executing a structural optimization task requires a deep understanding of optimization theory and often, programming skills. This complexity not only impedes the efficiency of design iterations but also limits the broader adoption of these advanced tools. Therefore, leveraging emerging AI technologies to simplify the interaction with these specialized tools, and to enable the intelligent definition and automated execution of analysis and optimization tasks, has become a pivotal research direction for enhancing the efficiency of modern engineering design.

To address this gap, this paper introduces a Multidisciplinary Design and Optimization Agent (MDO Agent) driven by LLMs. This framework establishes an interactive, semi-automated design system that spans from conceptualization to final optimization. By orchestrating a synergistic multi-agent system, the MDO Agent bridges the divide between unstructured natural language intent and formal engineering execution. This work provides a methodological foundation for the next generation of intelligent design systems, aiming to accelerate product innovation and enhance engineering efficiency. Our key contributions are:

\begin{itemize}
    \item Proposal of a synergistic multi-agent framework, the MDO Agent, which semi-automates the entire mechanical design workflow from natural language intent to a verified and optimized engineering solution.

    \item Systematic integration of tailored AI techniques for engineering design, including: Retrieval-Augmented Generation (RAG) for knowledge-grounded concept creation; a natural language-driven parametric modeling approach with visual self-correction for robustly generating complex CAD models; and a ReAct-style reasoning framework for the autonomous invocation of external engineering tools.

    \item Comprehensive validation of the MDO Agent's efficacy and robustness on complex, real-world engineering cases. The quantitative results demonstrate significant improvements in design efficiency, automation levels, and the potential to foster innovation.
\end{itemize}

The remainder of this paper is organized as follows. Section \ref{sec:proposed_method} elaborates on the proposed MDO Agent framework, detailing the architecture and core functionalities of its constituent agents. Section \ref{sec:experimental_evaluation} presents the experimental setup and provides in-depth analysis of the case studies, evaluating the performance of the MDO Agent across different design stages. Finally, Section \ref{sec:conclusion} concludes the paper and discusses promising directions for future research.


\section{MDO Agent Framework}
\label{sec:proposed_method}

\subsection{Overall Framework}
\label{subsec:overall_framework}

\begin{figure}[htbp]
    \centering
    \includegraphics[width=\textwidth]{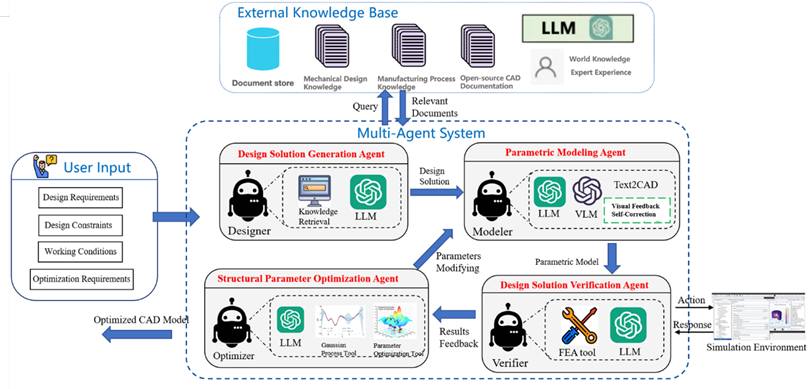}
    \caption{Schematic of the MDO Agent. The framework couples a Designer, Modeler, Verifier, and Optimizer in a closed loop, grounded by RAG for concept generation, text‑to‑CAD for parametric modeling, and ReAct‑style tool orchestration for FEA and optimization, with a human‑in‑the‑loop for guidance.}
    \label{fig:framework}
\end{figure}

This paper proposes a framework for a Large Language Model-driven Multidisciplinary Design and Optimization Agent, aiming to achieve an end-to-end intelligent design process from concept generation to parametric modeling, performance verification, and structural optimization. As illustrated in Figure \ref{fig:framework}, the core of this framework is a multi-agent system \cite{li2024survey} composed of four synergistic agents: the Designer Agent for design solution generation, the Modeler Agent for parametric modeling, the Verifier Agent for design validation, and the Optimizer Agent for structural parameter optimization.

The workflow of the MDO Agent closely aligns with typical mechanical design phases. The Designer agent, serving as the starting point, receives user design requirements and constraints expressed in natural language. By interacting with knowledge bases and leveraging the LLM's strong understanding and generation capabilities, it translates these into structured design solution text descriptions. Subsequently, the Modeler agent utilizes LLM and Vision-Language Model (VLM) Text-to-CAD capabilities to convert text descriptions into parametric CAD modeling code, automatically generating 3D models with visual feedback for self-correction to ensure accuracy. The Verifier agent integrates Finite Element Analysis (FEA) tool interfaces and employs LLM's reasoning abilities to automatically generate simulation code, execute analyses, and extract key performance indicators. Finally, the Optimizer agent integrates efficient optimization algorithms, collaborating with the Modeler and Verifier to optimize model parameters based on user requirements, thereby enhancing design performance.

Through the close collaboration and efficient information flow, the MDO Agent automates the entire process from understanding user needs, generating design concepts, constructing parametric models, performing performance verification, to ultimately achieving structural optimization. The system ultimately outputs the optimized CAD model, detailed optimization process, and final performance results, providing visual feedback to the user, thereby empowering efficient and intelligent mechanical product innovation design.

\subsection{Parametric Modeling with Natural Language Instructions}
\label{sec:nl_parametric_modeling}

\subsubsection{Generation, Editing, and Design Fusion}

\begin{figure}[htbp]
    \centering
    \includegraphics[width=\textwidth]{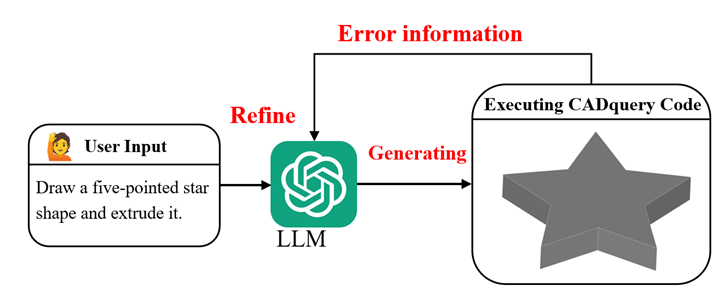}
    \caption{Text‑to‑geometry mapping. Natural‑language intent is translated into structured parametric code and geometric features to construct editable CAD models. }
    \label{fig:generating}
\end{figure}

The foundational capability of the Modeler Agent is the direct mapping from natural language to parametric CAD code. As depicted in Figure~\ref{fig:generating}, this process leverages the LLM's powerful semantic understanding and code generation abilities to convert a user's textual description (e.g., "draw a five-pointed star and extrude it") into an executable Python script using the CadQuery library. CadQuery is selected for its concise fluent API and robust parametric features. The initial generation process can be formalized as:
\begin{equation}
y_0 \sim P_{\text{LLM}}(y | x, E_g)
\label{eq:initial_generation}
\end{equation}
where $x$ is the user's natural language instruction, $y_0$ is the initially generated CAD script, $E_g$ represents few-shot examples provided to guide the LLM, and $P_{\text{LLM}}$ is the conditional probability distribution of the LLM. The generated script $y_0$ is then executed by a rendering engine to produce a 3D model, $M_0 = \text{Render}(y_0)$. If a code execution error occurs, the error message $F_{\text{err}}$ is fed back to the LLM to initiate a self-correction loop until executable code is produced.

\begin{figure}[htbp]
    \centering
    \includegraphics{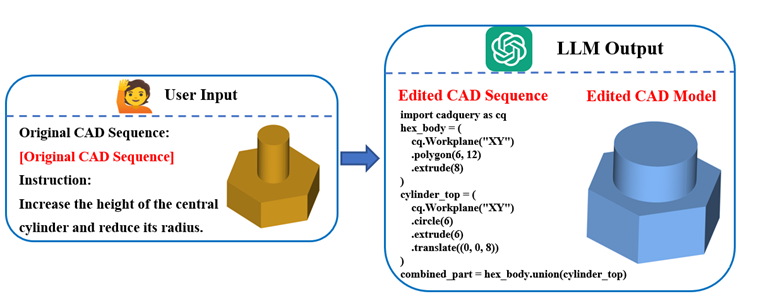}
    \caption{Interactive editing mechanism. The agent refines geometry through natural‑language edits that are compiled into code modifications and reapplied to the model.}
    \label{fig:interactive_editing}
\end{figure}

To support iterative refinement during the design process, we introduce an \textbf{interactive editing} mechanism. As shown in Figure~\ref{fig:interactive_editing}, a user can modify an existing model by providing a new natural language command $x'$. The process for the $k$-th iteration is described as:
\begin{equation}
y_k \sim P_{\text{LLM}}(y | x', y_{k-1}, E_g)
\label{eq:interactive_editing}
\end{equation}
where $y_{k-1}$ is the script from the previous step. This mechanism enables dynamic and fine-grained adjustments to the model's geometric parameters and topology.

\begin{figure}[htbp]
    \centering
    \includegraphics[width=\textwidth]{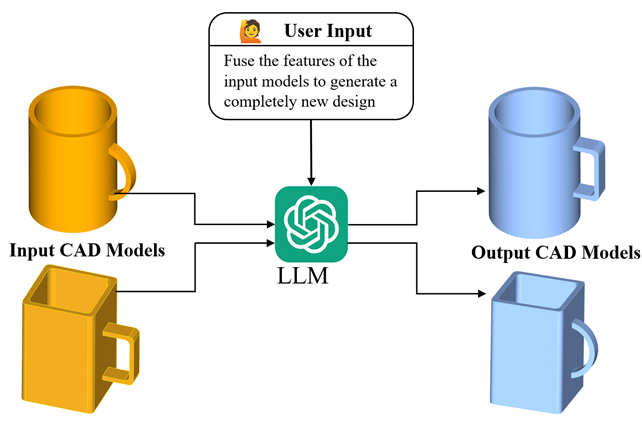}
    \caption{Design fusion mechanism. The agent merges elements from multiple parametric designs under natural‑language guidance, reconciling geometry and constraints to form a unified model.}
    \label{fig:design_fusion}
\end{figure}

Furthermore, to promote design reuse and innovation, we propose a \textbf{design fusion} mechanism (Figure~\ref{fig:design_fusion}). This allows the LLM to analyze the features of multiple existing design instances $\{D_1, D_2, \dots, D_n\}$ and intelligently combine them into a novel design solution $y_f$ based on a user's fusion command $x''$.

\subsubsection{Task Decomposition with Visual Self‑Correction}

Generating a complete modeling script for a complex mechanical structure in a single pass is challenging for an LLM due to its inherent limitations in spatial reasoning. To overcome this, we have devised an advanced modeling strategy that emulates the "divide and conquer" cognitive approach of human designers and integrates visual feedback for self-correction.

\begin{figure}[htbp]
    \centering
    \includegraphics[width=0.99\textwidth]{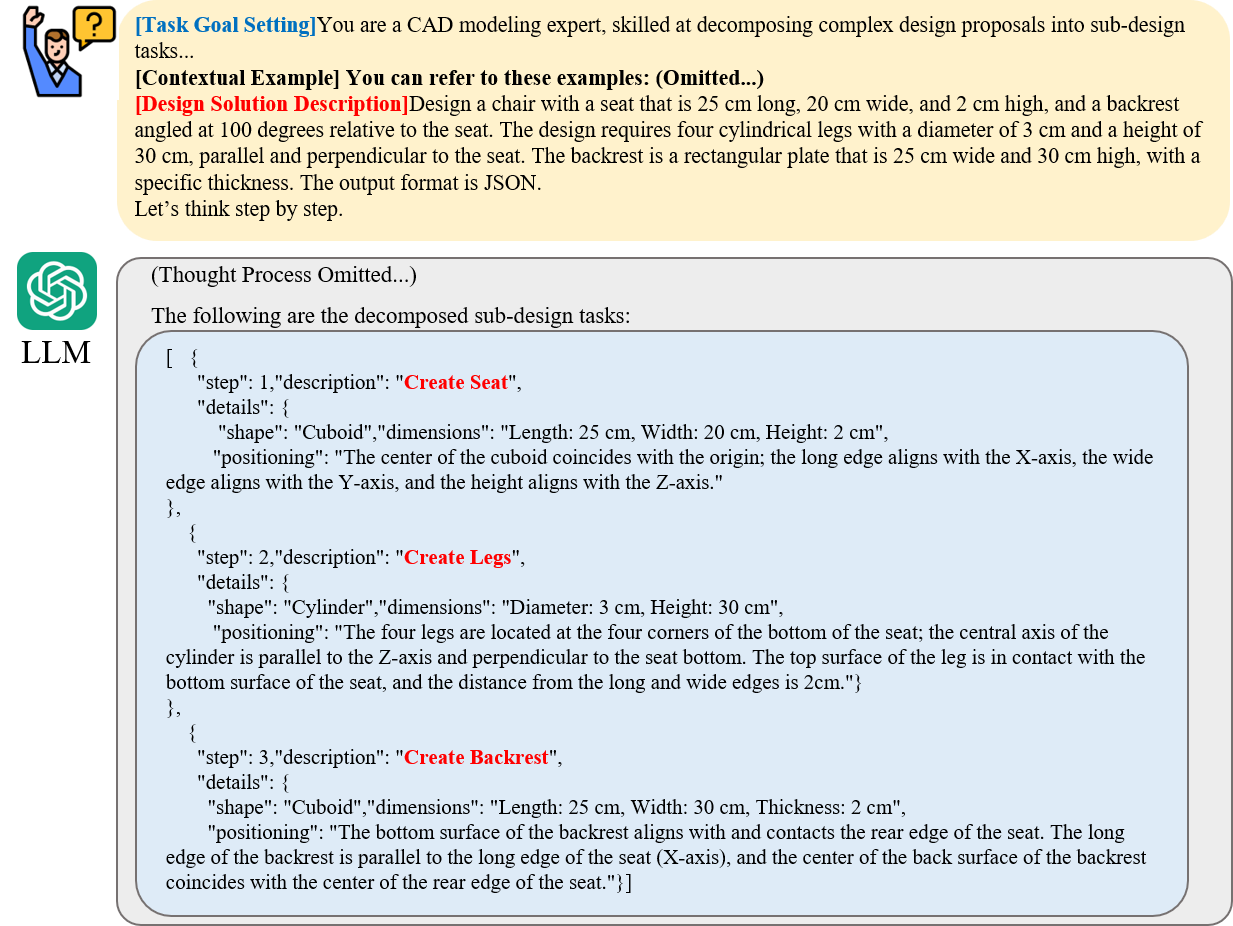}
    \caption{Chain‑of‑thought–based task decomposition. High‑level goals are broken into ordered sub‑tasks, each mapped to code generation or tool calls, enabling robust execution on complex geometry.}
    \label{fig:Task_Decomposition}
\end{figure}

\paragraph{Chain-of-Thought-based Task Decomposition} As shown in Figure~\ref{fig:Task_Decomposition}, the Modeler Agent first decomposes a complex modeling task into a sequence of logically independent subtasks. Using Chain-of-Thought reasoning, the LLM generates an ordered list of steps $\{T_1, T_2, \dots, T_N\}$. Each step contains not only the operational command but also precise spatial and geometric constraints, such as shape, dimensions, and positioning. For instance, designing a chair would be broken down into sub-tasks like creating the seat, adding the legs, and constructing the backrest.

\begin{figure}[htbp]
    \centering
    \includegraphics[width=\textwidth]{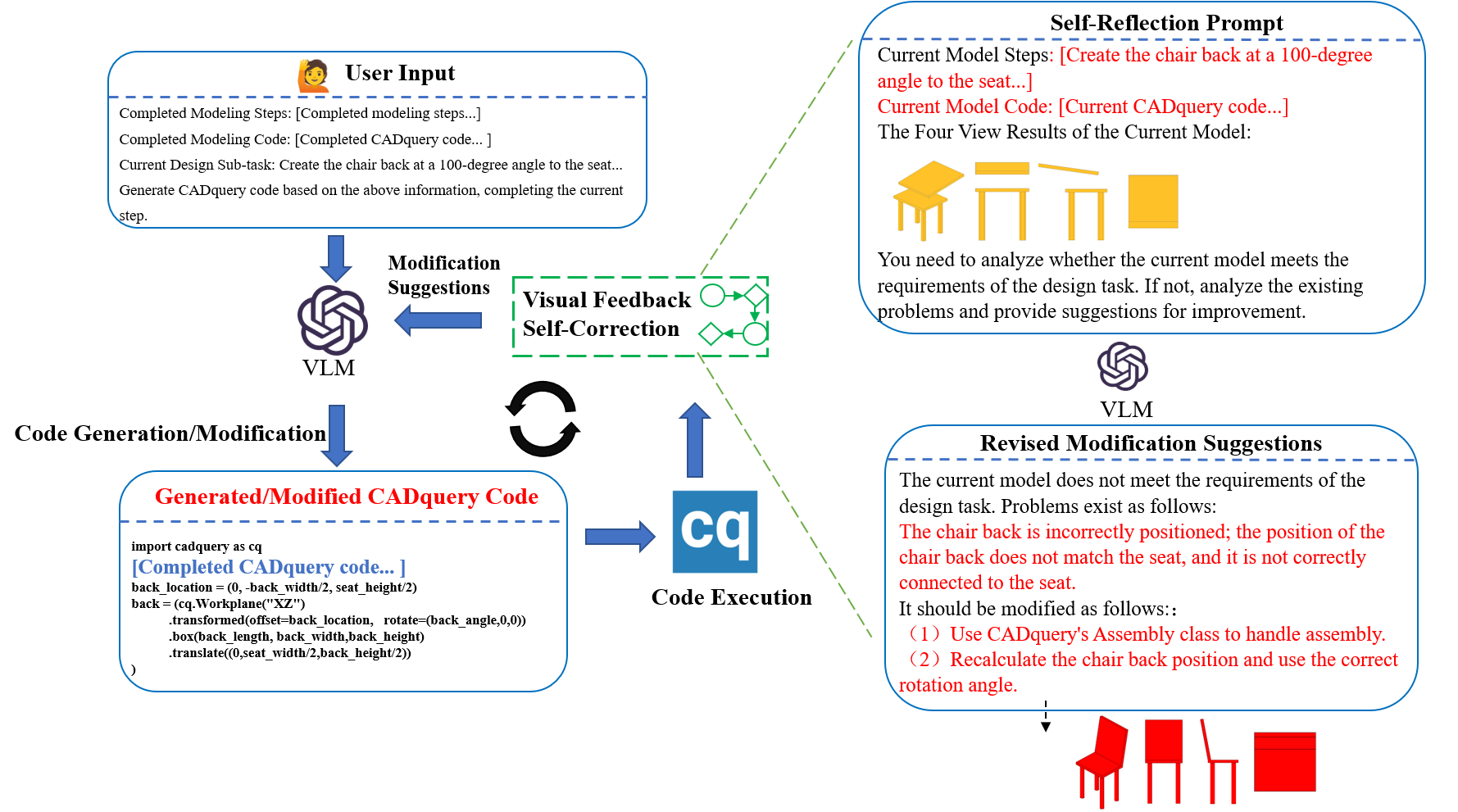}
    \caption{Visual feedback for self‑correction. A VLM inspects rendered results to detect geometric errors and guides iterative code fixes until the target intent is satisfied.}
    \label{fig:Visual_Feedback}
\end{figure}

\paragraph{Visual Feedback for Self-Correction} To ensure the accurate execution of each sub-task, we introduce a Vision-Language Model (VLM) \cite{zhang2024vision} in a closed-loop correction process, as illustrated in Figure~\ref{fig:Visual_Feedback}. This mechanism follows a "generate-render-correct" iterative cycle:
\begin{enumerate}
    \item \textbf{Generate:} The VLM generates the CAD code $y_i$ for the current subtask $T_i$.
    \item \textbf{Render:} The script $y_i$ is executed to produce multi-view renderings of the intermediate 3D model $M_i$.
    \item \textbf{Correct:} The VLM analyzes the rendered images and compares them against the sub-task description $T_i$. If discrepancies are found (e.g., dimensional errors, missing features), the VLM diagnoses the root cause and generates corrected code $y_{i+1}$.
\end{enumerate}

\begin{figure}[htbp]
    \centering
    \includegraphics[width=\textwidth]{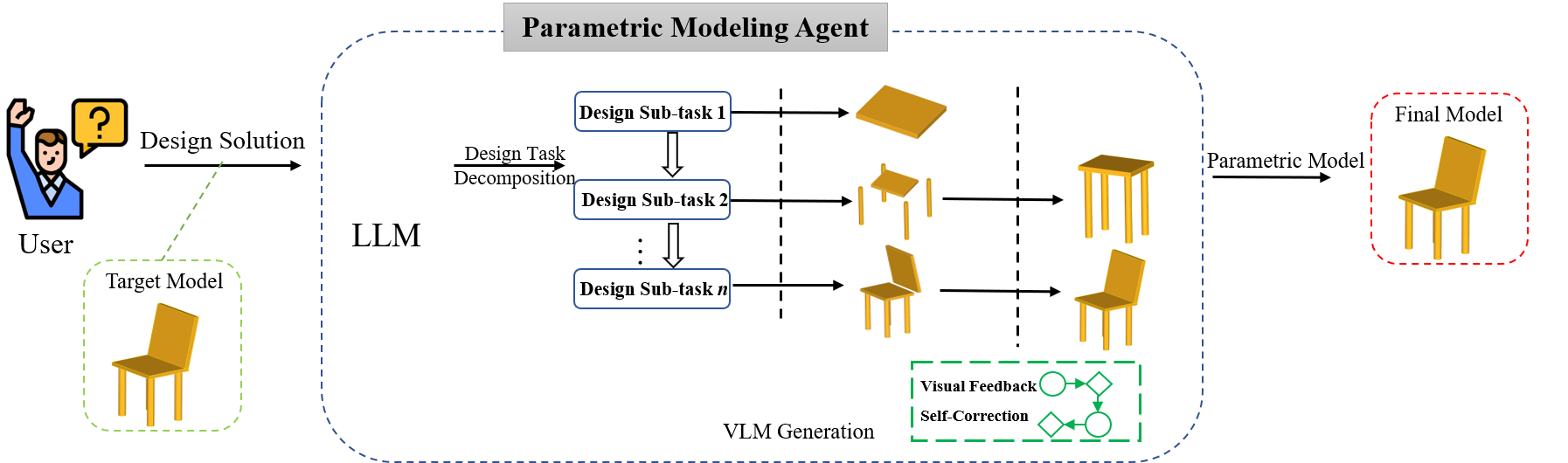}
    \caption{Pipeline of the Modeler Agent. The agent converts language instructions into CadQuery code, executes and renders the model, receives visual feedback, and iteratively refines the code.}
    \label{fig:Parametric_Modeling_Agent}
\end{figure}

This loop continues until the generated model fully aligns with the sub-task requirements or a maximum number of iterations is reached. In this manner, the agent can progressively and accurately construct complex mechanical structures. The overall principle of this advanced modeling agent is depicted in Figure~\ref{fig:Parametric_Modeling_Agent}.

\begin{figure}[htbp]
    \centering
    \includegraphics[width=\textwidth]{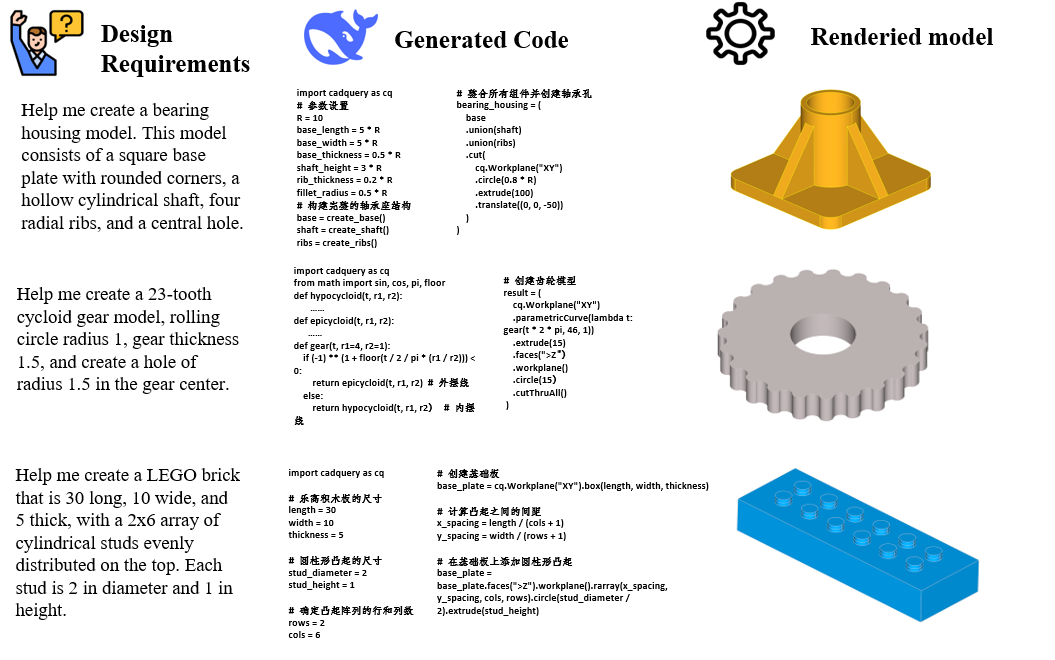}
    \caption{Example of language‑to‑parametric modeling. Representative results produced directly from natural‑language prompts illustrate the fidelity and editability of generated models.}
    \label{fig:parametric_modeling_example}
\end{figure}

\begin{figure}[htbp]
    \centering
    \includegraphics[width=\textwidth]{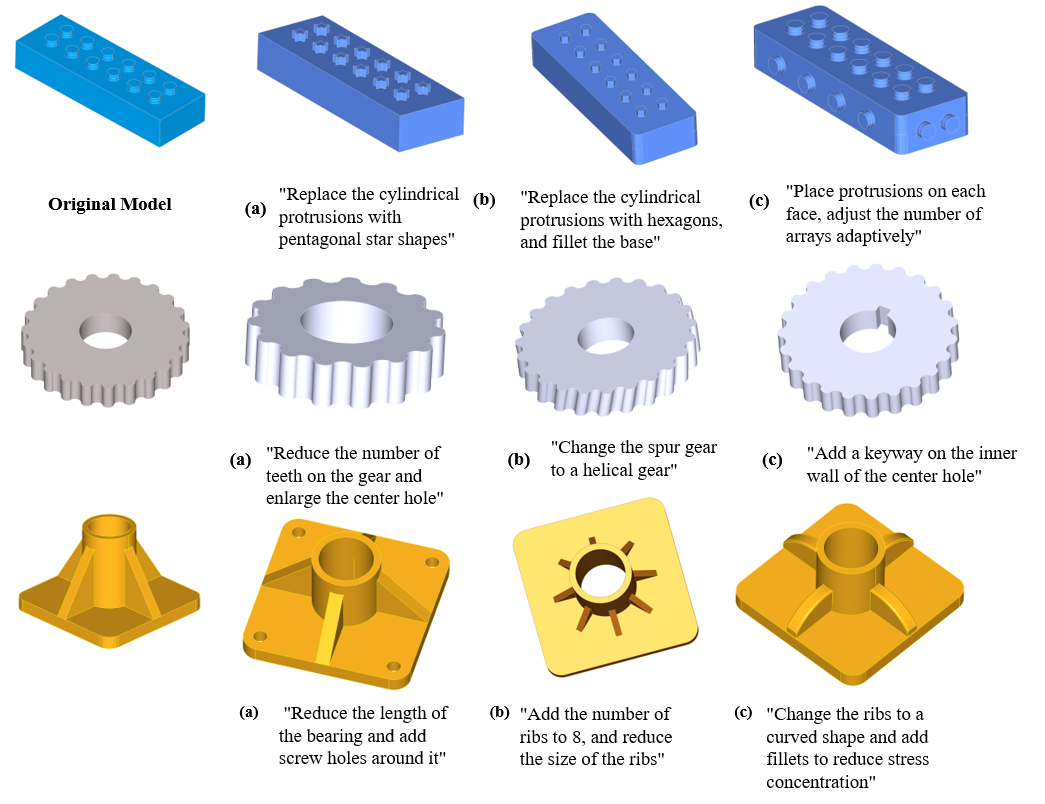}
    \caption{Example of interactive editing. Successive language edits adjust key features and constraints, demonstrating efficient human‑AI collaboration on model refinement.}
    \label{fig:interactive_editing_example}
\end{figure}

\begin{figure}[htbp]
    \centering
    \includegraphics[width=\textwidth]{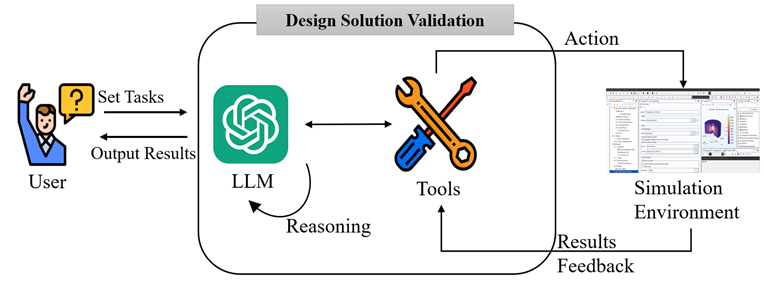}
    \caption{Example of multi‑design fusion. The agent composes features from multiple source designs into a cohesive parametric model under global constraints.}
    \label{fig:multi_design_fusion_example}
\end{figure}

Examples of parametric modeling, interactive editing, and multi-design fusion are shown in Figures \ref{fig:parametric_modeling_example}, \ref{fig:interactive_editing_example}, and \ref{fig:multi_design_fusion_example}, respectively.

\subsection{Conceptual Design Generation with Retrieval‑Augmented Generation}
\label{sec:rag_design_generation}

The primary role of the Designer Agent is to translate a user's natural language requirements into innovative and engineering-feasible initial design concepts. However, general-purpose LLMs face two significant shortcomings when applied to the specialized domain of mechanical engineering: their internal knowledge may be outdated or lack depth, and their inherent tendency for "hallucination" can produce outputs that violate physical laws or engineering common sense.

\begin{figure}[htbp]
    \centering
    \includegraphics[width=\textwidth]{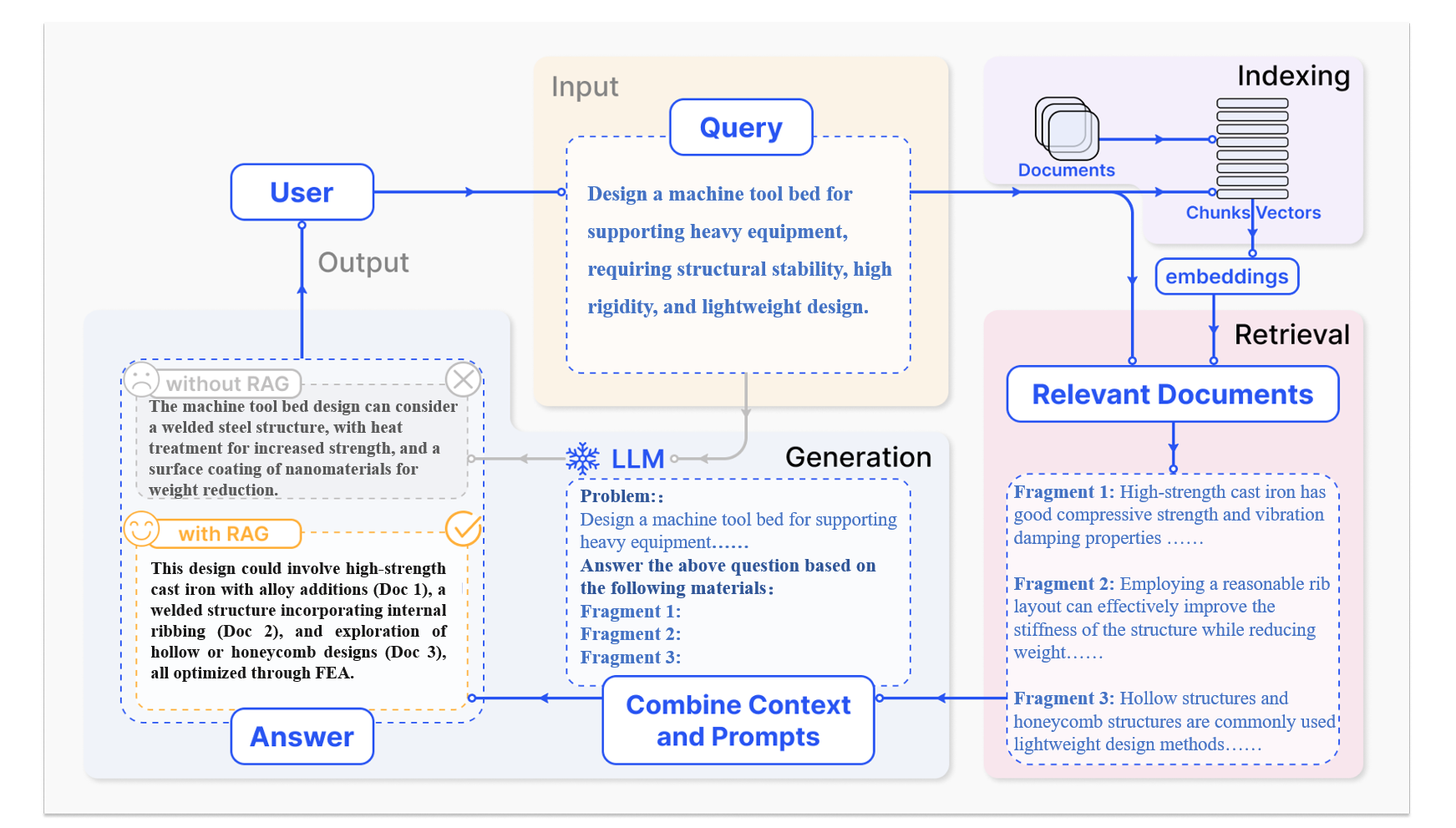}
    \caption{Retrieval‑Augmented Generation (RAG) pipeline. Domain knowledge is retrieved and fused with the prompt to ground concept generation and improve engineering feasibility.}
    \label{fig:rag_designer_agent}
\end{figure}

To overcome these limitations, we equip the Designer Agent with Retrieval-Augmented Generation technology \cite{koziolek2024llm}. As depicted in Figure~\ref{fig:rag_designer_agent}, the core mechanism of RAG is to ground the generative capabilities of the LLM in an external trusted knowledge base, thus ensuring the reliability and professionalism of the solutions generated \cite{xu2024retrieval}. The process comprises three main stages:

\paragraph{Indexing} We first construct a multi-source knowledge base containing authoritative design handbooks, engineering standards, and historical design cases. During the indexing stage, these documents are segmented into semantically coherent chunks. Each chunk, $c_i$, is then encoded into a high-dimensional vector $\mathbf{v}_i = \text{Embedding}(c_i)$ using a sentence-embedding model (e.g., bge-large-zh-v1.5 \cite{hu2024enhancing}). These vectors populate our vectorized knowledge base for subsequent retrieval.

\paragraph{Retrieval} When a user submits a design query, it is first encoded into a query vector $\mathbf{q} = \text{Embedding}(\text{Query})$ using the same embedding model. The system then computes the cosine similarity between the query vector $\mathbf{q}$ and every chunk vector $\mathbf{v}_i$ in the knowledge base:
\begin{equation}
\text{sim}(\mathbf{q}, \mathbf{v}_i) = \frac{\mathbf{q} \cdot \mathbf{v}_i}{\|\mathbf{q}\| \|\mathbf{v}_i\|}
\end{equation}
Based on the similarity scores, all chunks are ranked, and the top-$K$ most relevant chunks are selected to form a context set, $C^* = \{c_1^*, c_2^*, \dots, c_K^*\}$.

\paragraph{Generation} Finally, in the generation stage, the retrieved context set $C^*$ is integrated with the original user query into a structured prompt. This augmented prompt is then fed to the LLM to guide the generation of a design solution. Providing the LLM with relevant, factually accurate context significantly enhances the relevance and accuracy of its output. The generation process can be represented as:
\begin{equation}
\text{Response} \sim P_{\text{LLM}}(y | \text{Query}, C^*)
\end{equation}
To foster innovation and provide multiple options, we can employ different sampling strategies to generate several distinct candidate solutions for the designer to evaluate and select from.

By implementing the RAG mechanism, the Designer Agent can dynamically integrate external, verified domain knowledge into its reasoning process. This not only effectively mitigates the issues of knowledge gaps and outdated information in LLMs but, more critically, it substantially suppresses hallucinations by providing a factual basis for generation. Consequently, the engineering feasibility, logical consistency, and innovative potential of the generated design solutions are significantly enhanced.

\subsection{Optimization via Tool Calls and ReAct Orchestration}
\label{subsec:functional_optimization}

Traditional mechanical design optimization workflows are heavily reliant on manual programming to control complex analysis and optimization software, creating a significant bottleneck that hinders design iteration efficiency. To overcome this, our framework empowers agents with the ability to autonomously use external engineering tools through Large Language Model driven functional calls \cite{zhuang2023toolqa,wu2024avatar}. This transforms the LLM from a passive text generator into an intelligent agent that can plan, execute, and evaluate complex engineering tasks, enabling automated design verification and optimization.

The implementation of this capability involves two core components: a modularized engineering toolset and an autonomous reasoning-acting mechanism.

\begin{table}[htbp]
\centering
\caption{FEA toolset used by the Verifier and Optimizer Agents, including geometry import, meshing, boundary condition application, solving, and result extraction.}
\label{tab:fea_tools}
\begin{tabular}{llp{4cm}} 
\toprule
\textbf{Tool Name} & \textbf{Function Call} & \textbf{Description(Concise)} \\
\midrule
Geometry Loader & \texttt{load\_geometry()} & Loads and prepares the geometry file. \\
Physics Setup & \texttt{init\_physics\_field()} & Initializes the analysis physics (e.g., structural, thermal). \\
Entity Selector & \texttt{create\_entity\_set()} & Creates named geometric selections (faces, edges). \\
Material Definer & \texttt{define\_material\_properties()} & Defines material properties (e.g., elasticity, density). \\
Meshing Configurator & \texttt{set\_meshing\_options()} & Configures mesh settings like element type and size. \\
BC Applicator & \texttt{apply\_boundary\_conditions()} & Applies loads, constraints, and other boundary conditions. \\
Solver Executor & \texttt{execute\_solver()} & Executes the FEA solver to run the simulation. \\
Result Visualizer & \texttt{visualize\_results()} & Generates visualizations of the results (e.g., contour plots). \\
Code Interpreter & \texttt{execute\_python\_code()} & Runs custom Python code for advanced or complex tasks. \\
Human Handoff & \texttt{ask\_for\_human\_help()} & Requests human assistance for tasks beyond its capability. \\
\bottomrule
\end{tabular}
\end{table}

\paragraph{Modularized Engineering Toolset} We first construct a comprehensive, modular toolset by encapsulating discrete engineering operations into callable functions. This toolset comprises two main categories:

\begin{itemize}
    \item \textbf{FEA Tools:} As detailed in Table~\ref{tab:fea_tools}, we have wrapped a series of Finite Element Analysis operations into standardized tool functions. These tools cover the entire simulation pipeline, from model import and material definition to boundary condition application, solving, and results post-processing. This abstraction allows the LLM to perform complex simulations by composing a sequence of high-level function calls, rather than writing low-level code.
    
    \item \textbf{Optimization Tools:} The toolset is further augmented with a collection of powerful numerical optimization algorithms, such as genetic algorithms, particle swarm optimization, and Bayesian optimization \cite{wang2023recent}. These optimizers are also encapsulated as standardized components. The LLM can dynamically select, configure, and invoke the most appropriate optimizer based on the user's defined objectives, constraints, and the nature of the design variables.
\end{itemize}

\begin{figure}[htbp]
    \centering
    \includegraphics[width=\textwidth]{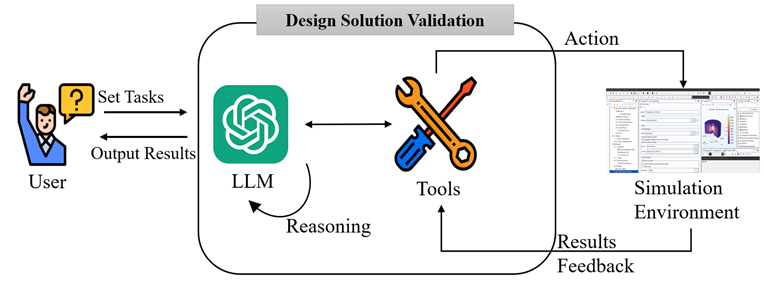}
    \caption{ReAct mechanism for autonomous tool use. Reasoning traces interleave with environment actions to complete FEA and form a closed loop with the optimizer.}
    \label{fig:react_workflow}
\end{figure}

\paragraph{Reasoning and Acting Mechanism} To enable the agent to intelligently decide which tool to use and when, we adopt a framework inspired by the Reasoning and Acting (ReAct) methodology \cite{yao2023react}. As illustrated in Figure~\ref{fig:react_workflow}, ReAct enables the LLM to synergize reasoning with external interaction, forming a "Thought-Action-Observation" loop. At each step $t$, the agent:
\begin{enumerate}
    \item \textbf{Thought:} Based on the current state $s_t$ (initialized with the user's task), the LLM generates an internal thought trace $T_t$, reasoning about the problem and planning the next action.
    \begin{equation}
        T_t = \text{LLM}_{\text{Think}}(s_t)
    \end{equation}
    \item \textbf{Action:} Based on its thought $T_t$, the LLM selects a tool to use and determines the appropriate arguments, forming an action $a_t$.
    \begin{equation}
        a_t = \text{LLM}_{\text{Act}}(T_t)
    \end{equation}
    \item \textbf{Observation:} The action $a_t$ is executed in the external environment (e.g., by calling an FEA tool), and the result is returned as an observation $o_t$.
\end{enumerate}
The observation is then incorporated back into the agent's state, $s_{t+1} = (s_t, T_t, a_t, o_t)$, and the loop continues until the task is completed.

Within our MDO Agent framework, the \textbf{Verifier Agent} leverages this ReAct mechanism to autonomously conduct performance analysis. Given a task like "evaluate the stress on this model under a 100N load," it will reason about the necessary steps and sequentially call the FEA tools to load the model, apply the boundary conditions, solve, and report the results. The \textbf{Optimizer Agent} represents a more advanced application of this paradigm. It orchestrates a higher-level loop, where it calls the optimization tools, which in turn invoke the Verifier Agent to evaluate the performance of each new design variant generated by the Modeler Agent. This creates a fully automated, closed-loop system for design optimization.

By combining a modular toolset with the ReAct reasoning framework, our agents can translate a user's high-level, natural language intent into a precise sequence of tool invocations. This achieves an unprecedented level of automation and intelligence in the design verification and optimization process, effectively bridging the gap between human intention and complex engineering execution.

\section{Experiments and Evaluation}
\label{sec:experimental_evaluation}

This section details the experimental setup and evaluation methodologies employed to comprehensively validate the proposed agent. We systematically assess its performance and practical value in key design stages: design solution generation, parametric modeling, and structural parameter optimization.

\subsection{Experimental Setup and Inputs}
\label{subsec:exp_setup_input}

The input data for our design tasks are structured around the following key elements:
\begin{itemize}
    \item \textbf{Design Requirements:} Specific functional goals, performance targets, and special product specifications.
    \item \textbf{Design Constraints:} Limitations on geometric dimensions, material properties, and operational conditions.
    \item \textbf{Operational Conditions:} Defined external environmental factors, including temperature, pressure, and applied loads.
    \item \textbf{Optimization Objectives:} Clearly stated goals, such as minimizing mass or maximizing strength, along with their associated constraint conditions.
\end{itemize}

Our experimental setup utilizes a multi-modal LLM capable of performing both Large Language Model and Vision-Language Model functions. Parametric CAD modeling is realized using the CadQuery library \cite{summers2006towards}, enabling seamless integration with the LLM. The Finite Element Analysis verification stage leverages the PyMechanical library in conjunction with commercial FEA software, Ansys Mechanical \cite{shaimi2022ansys}. Structural parameter optimization is implemented using the PyTorch deep learning framework \cite{imambi2021pytorch}.

To quantify the core capabilities of the MDO Agent at different design stages, we establish the key performance evaluation indicators as shown in Table~\ref{tab:kpis}.

\begin{table}[htbp]
    \centering
    \caption{Key performance indicators (KPIs) for evaluating the MDO Agent across concept generation, parametric modeling, verification, and optimization.}
    \label{tab:kpis}
    \resizebox{\textwidth}{!}{%
    \begin{tabular}{p{4cm} p{5cm} p{5cm}}
        \toprule
        \textbf{Design Stage} & \textbf{Evaluation Metric} & \textbf{Description} \\
        \midrule
        \textbf{Design Solution Generation} & Pass@N Rate & Proportion of valid concepts generated in N attempts, judged by experts. \\
        \addlinespace
        \textbf{Parametric Modeling} & Modeling Success & Binary outcome (Yes/No) indicating if a modeling sub-task is successfully completed within $\leq$ 3 iterations. \\
        & Generated Code Usability Rate & Percentage of agent-generated code that requires minimal or no manual editing for successful modeling. \\ 
        \addlinespace
        \textbf{Design Solution Verification (FEA)} & FEA Step Success & Binary outcome (Yes/No) indicating if an individual FEA setup/execution step is successfully completed from natural language instruction. \\
        & FEA Automation Contribution & Percentage of FEA workflow automated by agent's tool/API calls. \\
        \addlinespace
        \textbf{Structural Parameter Optimization} & Optimization Script Generation Success & Binary outcome (Yes/No) indicating if a correct, executable optimization script is generated in a single pass from natural language. \\
        & Generated Script Usability Rate & Percentage of agent-generated optimization script that requires minimal or no manual editing for successful execution. \\ 
        & Optimization Performance & Quantifiable improvement in objective function of optimized vs. baseline design. \\
        \bottomrule
    \end{tabular}%
    }
\end{table}

\subsection{Case Study I: Gas‑Turbine Blade}
\label{sec:case_turbine_blade}

\begin{figure}[h!]
    \centering
    \includegraphics[width=\textwidth]{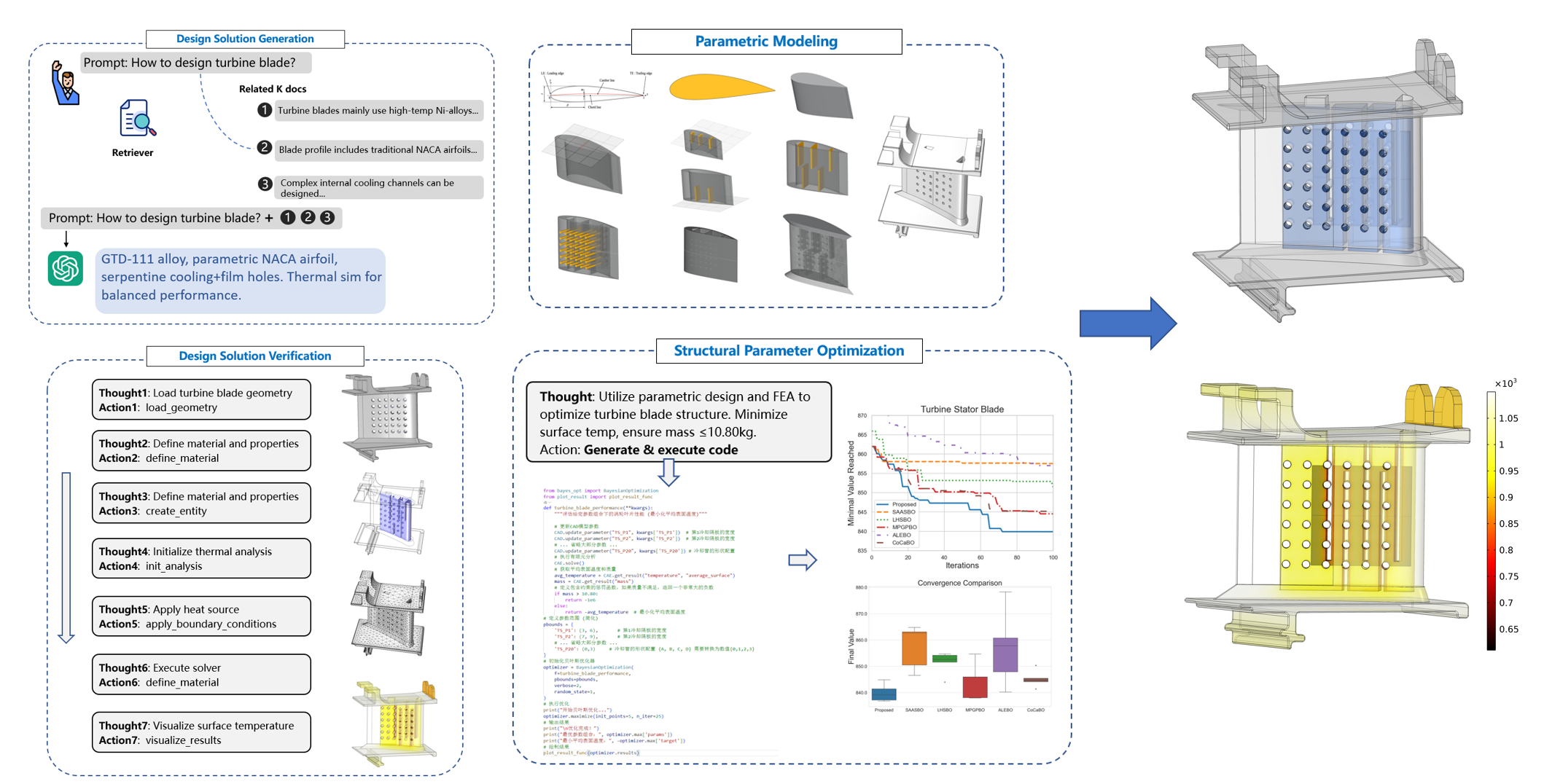}
    \caption{Gas turbine blade case. Problem setting, constraints, and operating conditions for the thermal‑structural design task.}
    \label{fig:turbine_blade_case}
\end{figure}

To comprehensively demonstrate and evaluate the integrated capabilities of the MDO Agent, we selected a gas turbine blade as our primary case study. This component is characterized by high geometric complexity and multi-physics challenges. In its operational environment, the blade must withstand pressures exceeding 40 bar and temperatures over 1000 K while enduring complex aerodynamic loads \cite{farahani2024power}. Thus, the design requires a delicate balance between structural integrity, cooling efficiency, and mass reduction. For this study, key design constraints include maintaining an average blade surface temperature below 840 K and a total mass under 10.80 kg. A visual representation is provided in Figure~\ref{fig:turbine_blade_case}.

\paragraph{Stage 1: Design Solution Generation}
The process begins with a user providing a natural language prompt, e.g., "Design a high-temperature, lightweight gas turbine blade for an inlet temperature of 1100 K, ensuring average surface temperature remains below 840 K and total mass is less than 10.80 kg." The Designer Agent then interprets this intent, retrieves relevant knowledge via RAG, and generates candidate solutions. Table~\ref{tab:turbine_designer_eval} presents five generated candidates and their validity assessment by domain experts. In repeated experiments, the \textit{Pass@5} rate reached 80\%, demonstrating the RAG mechanism's effectiveness in guiding the LLM towards concepts with high engineering feasibility.

\begin{table}[htbp]
\centering
\caption{Designer Agent–generated candidate solutions for the turbine blade and expert validity assessments.}
\label{tab:turbine_designer_eval}
\setstretch{1.3}
\resizebox{\textwidth}{!}{%
\begin{tabular}{lp{11cm}c}
\toprule
\textbf{No.} & \textbf{Solution Outline} & \textbf{Validity} \\
\midrule
1 & Utilizes GTD-111 directionally solidified nickel-based alloy. Airfoil based on a parameterized NACA 4-digit series. Features internal multi-pass serpentine cooling channels with turbulators and external film cooling holes. Requires thermo-mechanical coupled simulation for parameter optimization. & Yes \\
2 & Employs Inconel 718 alloy with an advanced CFD-optimized airfoil. Internal cooling combines impingement and pin-fin structures, with an external thermal barrier coating. Topology optimization is used for lightweighting, verified with FEA. & Yes \\
3 & Material is a tungsten-based alloy for extreme temperatures. Airfoil from an inverse design method. Internal cooling omitted in favor of high-efficiency external spray cooling and special surface micro-channels. & No \\
4 & Uses CMSX-4 single-crystal superalloy. Airfoil is an enhancement of a mature gas turbine model. Features a complex internal tree-like cooling network and full-coverage effusion cooling holes. A multi-objective genetic algorithm optimizes parameters. & Yes \\
5 & Blade body made of Ceramic Matrix Composites (CMC). Thin, high-lift airfoil design. Internal cooling uses micro-channel evaporative technology. Focus on CMC joining and sealing. & Yes \\
\bottomrule
\end{tabular}}
\end{table}

\begin{table}[htbp]
\centering
\caption{Parametric modeling performance on the turbine blade: sub‑task success within $\leq$3 iterations and usability of generated code.}
\label{tab:turbine_modeler_eval}
\begin{tabular}{p{4cm} p{3cm} p{3cm}p{4cm}} 
\toprule
\textbf{Modeling Sub-task} & \textbf{Modeling Success (within $\leq$ 3 iter.)} & \textbf{Generated Code Usability Rate (\%)} & \textbf{Remarks} \\ 
\midrule
Airfoil Profile Generation& Yes & 100 & Code generated with high direct usability. \\ 
Hollow Airfoil Creation & Yes & 100 & Automated Boolean operations, highly usable code. \\ 
Internal Rib Generation & Yes & 60 & Parametric generation successful (Yes), but required some manual adjustments for optimal placement. \\ 
Film Hole Generation & No & 65 & Successful parametric control (Yes), some fine-tuning needed due to complex surface projection. \\ 
Root Structure Generation & Yes & 90 & Automated creation resulted in highly usable code. \\ 
\midrule
\textbf{Average} & \textbf{80\% (Success Rate)} & \textbf{83.6\% (Average Usability Rate)} & \textbf{Summary of overall performance.} \\ 
\bottomrule
\end{tabular}
\end{table}

\paragraph{Stage 2: Parametric Modeling}
Once a preliminary design is selected, the \textbf{Modeler Agent} translates it into a precise, parametric CAD model. Given the blade's geometric complexity, the agent decomposes the task into sub-tasks: airfoil profile generation \cite{khater2023effectives}, hollow structure creation, internal turbulator placement, film cooling hole generation, and root structure construction. For each sub-task, the agent generates CadQuery code, leveraging VLM-based visual feedback for self-correction. The final model incorporates a complex internal cooling passage and a precise layout of film cooling holes. Table~\ref{tab:turbine_modeler_eval} quantifies the agent's efficiency, showing an average success rate of 80\% and an average code reduction of 83.6\% compared to manual scripting.

\paragraph{Stage 3: Verification and Optimization}
In the verification stage, the user submits a natural language request to the \textbf{Verifier Agent}, e.g., "Perform a steady-state thermal analysis on the current blade model and retrieve key temperature data." The agent activates its ReAct workflow, sequentially invoking FEA tools to load the geometry, define materials, apply boundary conditions (with potential human assistance for complex region selection), mesh the model, execute the solver, and extract results. The \textbf{Optimizer Agent} then takes over, responding to goals like "minimize the average blade surface temperature." It automatically generates the optimization script, forming a closed loop with the Modeler and Verifier to iteratively seek the optimal design.

\begin{table}[htbp]
\centering
\caption{Verification and optimization performance for the turbine blade: automated FEA steps, automation contribution to the workflow, and optimization script generation/execution outcomes.}
\label{tab:turbine_fea_opt_eval}
\resizebox{\textwidth}{!}{%
\begin{tabular}{p{4cm} p{3cm} p{3cm} p{4cm}} 
\toprule
\textbf{FEA / Optimization Sub-task} & \textbf{FEA Step Success (Y/N)} & \textbf{Automation Contribution (\%)} & \textbf{Remarks} \\ 
\midrule
Model Import \& Preprocessing & Yes & 100\% & API calls for geometry loading and checking accurately generated. \\
Material \& Physics Setup & Yes & 100\% & Agent automatically generates commands for material properties and physics setup. \\
Named Entity Selection & \textbf{No} & \textbf{10\%} & \textbf{Agent struggled with complex geometric entities, requiring manual selection of critical regions for boundary conditions.} \\ 
Boundary Condition Application & Yes & 50\% & Agent generates code, but precise selection on complex geometry may require human aid. \\
Meshing \& Solver Setup & Yes & 100\% & Generates meshing parameters and solver configurations from simple instructions. \\
\midrule
\textbf{Overall FEA Automation} & No & \textbf{70\%} & \\ 
\midrule
Optimization Script Generation & Yes & 100\% & Accurately translates optimization goals into a complete, executable script. \\
\bottomrule
\end{tabular}}
\end{table}

Table~\ref{tab:turbine_fea_opt_eval} summarizes the MDO Agent's performance in this stage. The agent automated approximately 70\% of the FEA workflow, and the optimization code generation success rate was 100\%, demonstrating its ability to accurately interpret user intent and orchestrate the entire optimization process.

\paragraph{Case Study I Summary}
The gas turbine blade case study demonstrates the significant potential of the MDO Agent. The introduction of RAG enhanced the engineering feasibility of initial concepts. The natural language-driven modeling process automated over 80\% of the modeling tasks and reduced coding effort by over 83.6\%. In the verification and optimization stages, the agent automated approximately 70\% of the FEA workflow and achieved fully autonomous execution of the optimization loop. These quantitative results confirm the MDO Agent's profound advantages in improving design efficiency and lowering technical barriers. While human assistance is still beneficial for highly complex geometric interactions, the framework establishes a solid foundation for a higher level of design intelligence.

\subsection{Case Study II: Machine‑Tool Column}
\label{sec:case_machine_tool}

\begin{figure}[h!]
    \centering
    \includegraphics[width=\textwidth]{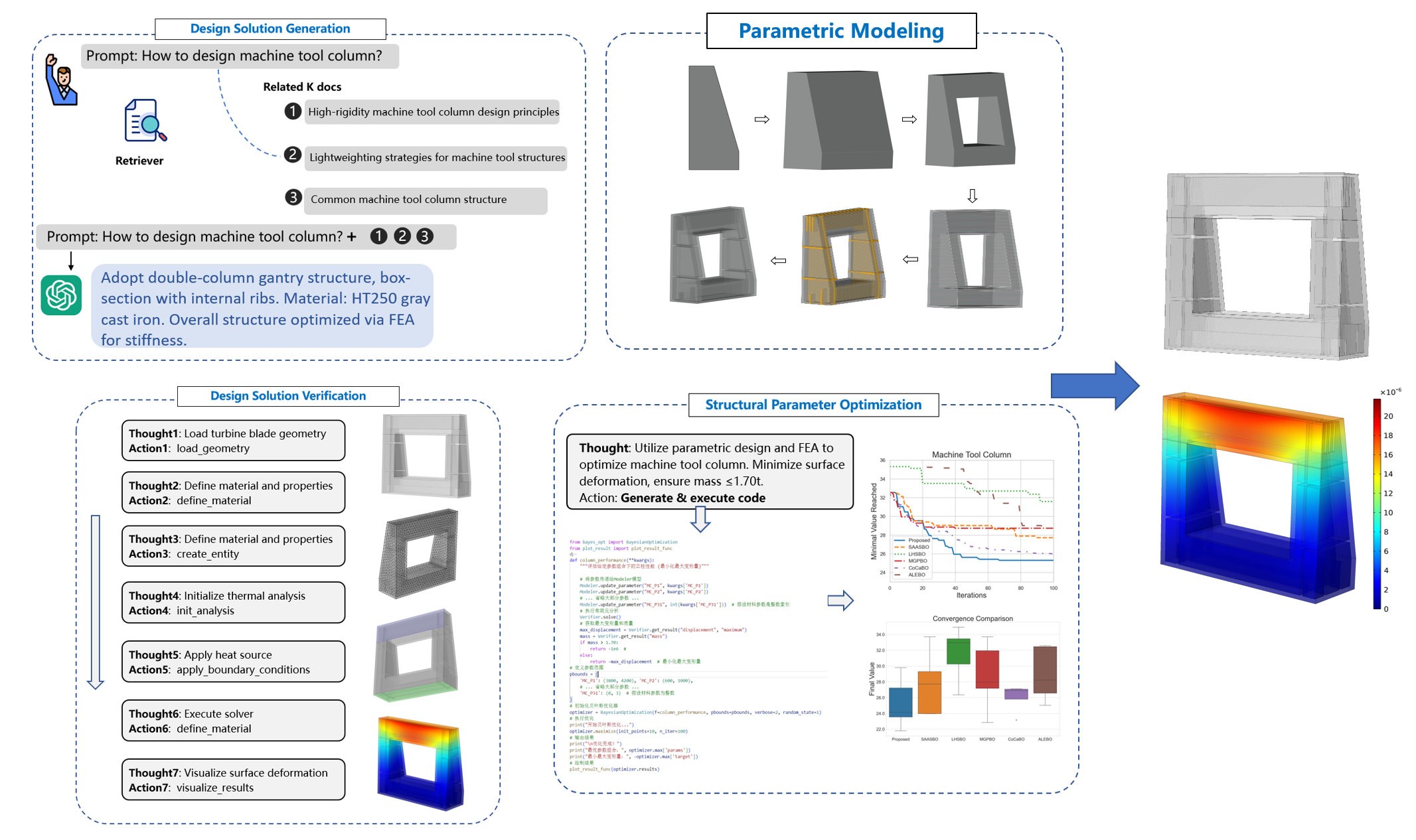}
    \caption{Machine tool column case. Geometry, loading conditions, and constraints used to evaluate stiffness and weight trade‑offs.}
    \label{fig:machine_tool_case}
\end{figure}

To further validate the MDO Agent's versatility, we select the column of a five-axis machine tool as a second case study. As a critical load-bearing and support component, the structural rigidity, dynamic characteristics, and thermal stability of the column profoundly impact the overall performance of the machine \cite{venugopal2019structural}. The design must balance high stiffness with lightweighting to achieve an optimal trade-off between performance, energy efficiency, and cost. Key design constraints include a total mass not exceeding 1.70 tonnes and a maximum global deformation of less than 25 \textmu m under a nominal 3 kN axial load. The case is depicted in Figure~\ref{fig:machine_tool_case}.

\paragraph{Stage 1: Design Solution Generation}
The \textbf{Designer Agent} initiates the process by generating preliminary design concepts based on a user's natural language prompt, e.g., "Design a high-stiffness, lightweight column for a five-axis machine tool, with a mass under 1.70 t and max deformation below 25 \textmu m under a 3 kN load." The agent's RAG mechanism retrieves relevant domain knowledge to ground the LLM's output. As shown in Table~\ref{tab:machinetool_designer_eval}, the agent produced five distinct concepts. The effective generation rate (\textit{Pass@5}) again reached 60\%, reaffirming the RAG's capability to generate viable engineering solutions.

\begin{table}[htbp] 
    \centering
    \caption{Designer Agent–generated preliminary solutions for the machine tool column and expert assessments.}
    \label{tab:machinetool_designer_eval} 
    \begin{tabularx}{\textwidth}{l X l} 
        \toprule
        \textbf{No.} & \textbf{Solution Outline} & \textbf{Validity} \\
        \midrule
        1 & Double-column gantry structure with rigid top beam. Trapezoidal cross-section columns with internal ribs, HT250 gray cast iron. & Yes \\
        2 & Frame structure, closed-box main member. Internal honeycomb, welded steel, FEA-optimized. & Yes \\
        3 & High-strength aluminum profiles, damping material fill. Focus on connection stiffness via pre-tightened bolts. & No \\
        4 & Monolithic casting with complex internal cavities/ribs for rigidity and mass reduction. Ductile iron, heat-treated. & Yes \\
        5 & Carbon fiber composite column, lightweight thin-walled structure. Focus on material choice and connection reliability, FEA-verified. & No \\
        \bottomrule
    \end{tabularx}
\end{table}

\paragraph{Stage 2: Parametric Modeling}
Following concept selection, the \textbf{Modeler Agent} converts the textual description into a parametric CAD model. The task is decomposed into sub-tasks: generating the trapezoidal support base, creating rectangular openings via Boolean operations, shelling the main body, and parametrically arranging the internal ribs. The resulting parametric model is shown in Figure~\ref{fig:machine_tool_case}, with its design variables detailed in the study. As quantified in Table~\ref{tab:machinetool_modeler_eval}, the agent achieved an impressive average success rate of 75\% and an average code reduction of 76.2\%, highlighting its efficiency and reliability in automating the modeling process.

\begin{table}[htbp]
\centering
\caption{Parametric modeling performance on the machine tool column: modeling success within $\leq$3 iterations and code usability.} 
\label{tab:machinetool_modeler_eval}
\begin{tabular}{p{3cm} p{2.5cm} p{2.5cm}p{4cm}} 
\toprule
\textbf{Modeling Sub-task} & \textbf{Modeling Success (within $\leq$ 3 iter.)} & \textbf{Generated Code Usability Rate (\%)} & \textbf{Remarks} \\ 
\midrule
Trapezoidal Base Generation & Yes & 80 & 2D profile and extrusion successfully automated. \\
Rectangular Cutout & Yes & 85 & Boolean operations and positioning automated. \\
Column Shelling & Yes & 100 & Shelling operation successfully automated. \\
Internal Rib Layout & No & 40 & Difficulties with spatial reasoning led to errors; required significant manual correction. \\ 
\midrule
\textbf{Average} & \textbf{75\% (Success Rate)} & \textbf{76.2\% (Average Usability Rate)} & \textbf{Summary of overall performance.} \\ 
\bottomrule
\end{tabular}
\end{table}

\begin{table}[htbp]
\centering
\caption{Machine tool column—verification and optimization outcomes, including FEA automation contribution and optimization script usability.}
\label{tab:machinetool_fea_opt_eval}
\resizebox{\textwidth}{!}{%
\begin{tabular}{p{3.5cm} p{2.5cm} p{2.5cm} p{3.5cm}} 
\toprule
\textbf{FEA / Optimization Sub-task} & \textbf{FEA Step Success (Y/N)} & \textbf{Automation Contribution (\%)} & \textbf{Remarks} \\ 
\midrule
Model Import \& Preprocessing & Yes & 100\% & Accurately generated API calls for STEP file loading and initial checks. \\
Material \& Physics Setup & Yes & 100\% & Automatically configured material properties (HT250) and analysis type. \\
\textbf{Named Entity Selection} & \textbf{No} & \textbf{40\%} & \textbf{Agent struggled with complex geometric entities, requiring manual selection of critical regions for boundary conditions.} \\ 
Boundary Condition Application & Yes & 70\% & Effectively handled standard constraints and loads on well-defined faces. \\ 
Meshing \& Solver Setup & Yes & 100\% & Generated meshing and solver settings from simple user commands. \\
\midrule
\textbf{Overall FEA Automation} & No & \textbf{82\%} & \\ 
\midrule
Optimization Script Generation & Yes & 100\% & Accurately generated a complete, executable Python script for the optimization loop. \\
\bottomrule
\end{tabular}}
\end{table}

\paragraph{Stage 3: Verification and Optimization}
The user then tasks the \textbf{Verifier Agent} with a natural language command: "Set up a static structural analysis for the column. Apply a 3 kN downward load in the Z-direction on the front face of the crossbeam and a fixed support on the bottom face. Solve for stress and strain." The agent activates its ReAct workflow, autonomously calling FEA tools to load the model, define the material (HT250), apply boundary conditions, mesh, and solve, reporting a maximum deformation of 30.79 \textmu m. Table~\ref{tab:machinetool_fea_opt_eval} details the agent's performance, showing a high degree of automation, particularly for this geometrically regular structure. The \textbf{Optimizer Agent} subsequently uses this verified workflow within its optimization loop to intelligently tune structural parameters, aiming to meet all design constraints.

\paragraph{Case Study II Summary}
The machine tool column case further validates the MDO Agent's broad applicability and effectiveness. The RAG-powered design generation provided a strong conceptual foundation. The natural language-driven modeling process again proved highly efficient, significantly reducing manual effort. In the verification and optimization phase, the agent demonstrated robust tool-use capabilities, automating a large portion of the standard FEA workflow and seamlessly setting up the optimization task. The results collectively affirm that the MDO Agent serves as a powerful tool for accelerating the design cycle of complex yet geometrically regular structures.

\subsection{Case Study III: Fractal Heat Sink}
\label{sec:case_heat_sink}

\begin{figure}[h!]
    \centering
    \includegraphics[width=\textwidth]{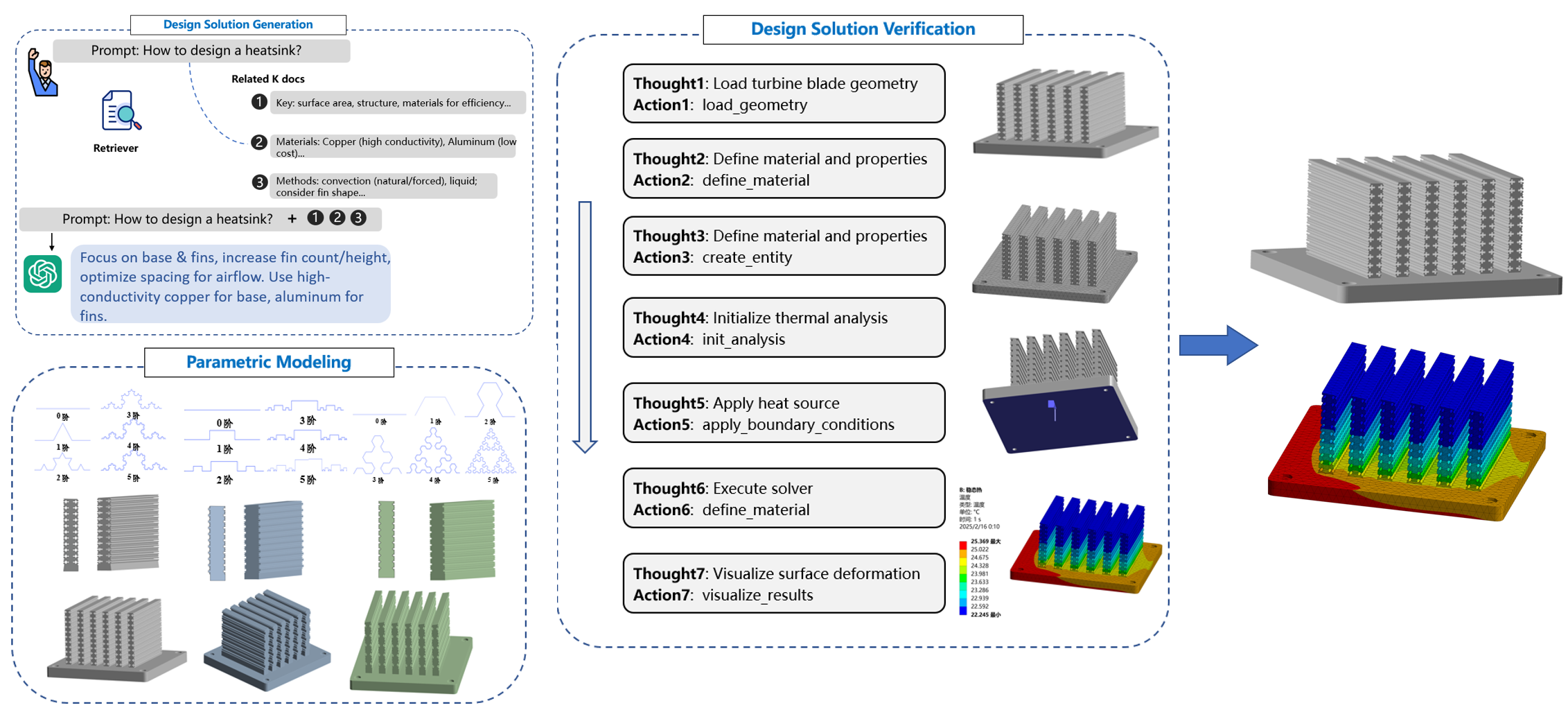} 
    \caption{Fractal heat sink case. Koch‑like fractal channels are parameterized for thermal–hydraulic evaluation and optimization.}
    \label{fig:heat_sink_case}
\end{figure}

To further test the MDO Agent's capacity for handling components with complex geometric features and specific physical performance requirements, we select a fractal heat sink as our third case study. As the power density of electronic devices increases, conventional heat sink designs face significant thermal dissipation bottlenecks. Fractal geometries, with their unique self-similarity and high surface-area-to-volume ratios, offer tremendous potential for enhancing heat transfer \cite{vu2022fractal}. This case aims to evaluate the agent's ability to generate and model such non-conventional, algorithmically-defined structures. The design is depicted in Figure~\ref{fig:heat_sink_case}.

\begin{table}[htbp]
\centering
\caption{Fractal heat sink—generated designs and expert assessments.}
\label{tab:heatsink_designer_eval}
\resizebox{\textwidth}{!}{%
\begin{tabular}{lp{11cm}c}
\toprule
\textbf{No.} & \textbf{Solution Outline} & \textbf{Validity} \\
\midrule
1 & Copper base plate with aluminum fins. The fin edges are designed with a Koch fractal structure. & Yes \\
2 & Copper base plate with aluminum fins. The fin edges are designed with a Sierpinski arrowhead curve fractal. & Yes \\
3 & Copper base plate with aluminum fins. The fin edges are designed with a Cantor set fractal structure. & Yes \\
4 & Utilizes a vapor chamber combined with high-conductivity materials and natural convection fins for a compact, high-efficiency design. & Yes \\
5 & Employs heat pipe technology to conduct heat to external fins, cooled by natural or forced convection. & Yes \\
\bottomrule
\end{tabular}}
\end{table}

\begin{table}[htbp]
\centering
\caption{Parametric modeling performance for the fractal heat sink: success rate within $\leq$3 iterations and code usability.} 
\label{tab:heatsink_modeler_eval}
\begin{tabular}{p{4cm} p{2cm} p{2cm}p{3cm}} 
\toprule
\textbf{Modeling Sub-task} & \textbf{Modeling Success (within $\leq$ 3 iter.)} & \textbf{Generated Code Usability Rate (\%)} & \textbf{Remarks} \\ 
\midrule
Fractal Curve Definition & Yes & 70 & Successfully generated the recursive function. \\ 
Fin Profile & Yes & 100 & Created fin geometry based on the parameterized curve. \\ 
Fin Array Patterning & Yes & 100 & Automated array creation. \\ 
Base Plate Creation & Yes & 100 & Automatically draws the base plate and adds fillets. \\ 
Assembly & Yes & 85 & Defined constraints for geometric and assembly validity. \\ 
\midrule
\textbf{Average} & Yes & 91\%  & \textbf{Summary of overall performance.} \\ 
\bottomrule
\end{tabular}
\end{table}

\paragraph{Stage 1: Design Solution Generation}
The process starts with a relatively open-ended user prompt to the \textbf{Designer Agent}: "Help me design a heat sink for air convection cooling, consisting of a base plate and fins. Improve the structure to enhance cooling efficiency." Leveraging its RAG-enhanced knowledge base, the agent successfully retrieved information on advanced cooling techniques, including fractal geometries \cite{wang2023performance}. As shown in Table~\ref{tab:heatsink_designer_eval}, it generated five candidate solutions, four of which incorporated different fractal patterns. The effective generation rate (\textit{Pass@5}) reached 100\%, demonstrating the agent's powerful capability to access specialized knowledge and propose innovative, high-performance design concepts.

\paragraph{Stage 2: Parametric Modeling}
Upon selecting a fractal design, the \textbf{Modeler Agent} undertakes the task of creating the parametric model. This case highlights the agent's strength in handling algorithmically-defined geometry. The modeling is decomposed into sub-tasks: defining the fractal curve function, creating and extruding the fractal fin profile, patterning the fins in an array, and creating the base plate. As detailed in Table~\ref{tab:heatsink_modeler_eval}, the agent achieved a 100\% success rate across all sub-tasks and an average code reduction of 78\%. This exceptional performance underscores the agent's significant advantage in automating the creation of complex, programmatically generated geometries, a task that is tedious and error-prone when done manually.

\paragraph{Stage 3: Verification}
In the final stage, the \textbf{Verifier Agent} is tasked with performing a steady-state thermal analysis. The agent's ReAct loop coordinates the invocation of the FEA toolset APIs to perform model import, material definition, boundary condition setup (e.g., applying an 80 W heat source at the base), meshing, and solving. As shown in Table~\ref{tab:heatsink_fea_opt_eval}, the agent demonstrated a high degree of automation, with most steps achieving a 90-100\% first-pass success rate. The overall automation contribution was approximately 78\%, confirming the agent's ability to handle different physics (thermal analysis) with the same robust tool-use mechanism.

\begin{table}[htbp]
\centering
\caption{Fractal heat sink—automated verification metrics by FEA step, overall FEA automation contribution, and optimization script outcomes.}
\label{tab:heatsink_fea_opt_eval}
\resizebox{\textwidth}{!}{%
\begin{tabular}{p{4cm} p{3.5cm} p{3.5cm} p{4cm}} 
\toprule
\textbf{FEA Analysis Step} & \textbf{FEA Step Success (Y/N)} & \textbf{Automation Contribution (\%)} & \textbf{Remarks} \\ 
\midrule
Model Import \& Preprocessing & Yes & 100\% & Successfully automated model loading via \texttt{load\_geometry} function. \\
Material \& Physics Setup & Yes & 100\% & Automatically defined material properties and convection boundary conditions. \\
Named Entity Selection & No & 10\% & Struggled with complex fractal geometry; manual selection of critical regions for boundary conditions was necessary. \\ 
Boundary Condition Application & Yes & 80\% & Handled heat source (80 W) application to the base plate. \\
Meshing \& Solver Setup & Yes & 100\% & Demonstrated high automation in meshing and solver configuration. \\
\midrule
\textbf{Overall FEA Automation} & \textbf{N/A} & \textbf{78\%} & \textbf{High overall automation, with challenges in precise entity selection.} \\ 
\bottomrule
\end{tabular}}
\end{table}

\paragraph{Case Study III Summary}
The fractal heat sink case study effectively showcases the MDO Agent's ability to drive innovation by handling complex, algorithmically-defined geometries. The agent not only generated novel design concepts but also demonstrated exceptional efficiency in modeling them, a traditional pain point in CAD. The successful automation of the thermal analysis workflow further proves the versatility of the tool-use framework. This case provides compelling evidence that the MDO Agent is a powerful asset for exploring and realizing innovative solutions in complex engineering problems.

\subsection{Discussion}
\label{sec:results_discussion}

The experimental results from the three case studies—a gas turbine blade, a machine tool column, and a fractal heat sink—collectively validate the effectiveness and versatility of the MDO Agent. This section discusses the key findings and current limitations of the proposed framework.

Across all case studies, the MDO Agent demonstrated a robust and consistent workflow, successfully automating the design process from natural language input to a final, optimized solution. The framework's primary strength lies in its significant enhancement of design automation and efficiency. Quantitatively, the agent consistently reduced the parametric modeling workload by over 70\% (in terms of code) and automated approximately 70-75\% of the FEA simulation process (see Table~\ref{tab:results_summary}). This was achieved by seamlessly integrating design, modeling, and analysis tools through LLM-driven orchestration, which not only accelerates the design cycle but also lowers the technical barrier for using advanced engineering software. Furthermore, the fractal heat sink case highlighted the agent's potential to foster innovation by effectively handling non-conventional, algorithmically-defined geometries.

\begin{table}[htbp]
\centering
\caption{Overall quantitative comparison across the three case studies, aggregating modeling, verification, and optimization metrics.}
\label{tab:results_summary}
\resizebox{\textwidth}{!}
{%
\begin{tabular}{p{3cm} p{2cm} p{2cm} p{2cm} p{2cm} p{5cm}}
\toprule
\textbf{Case Study} & \textbf{Generation (Pass@5)} & \textbf{Modeling (Code Reduction)} & \textbf{FEA Automation} & \textbf{Opt. Code Gen.} & \textbf{Key Finding} \\
\midrule
Turbine Blade & 80\% & 83.6\% & 70\% & 100\% & Effective for complex multi-physics problems. \\
Machine Tool Column & 60\% & 76.2\% & 82\% & 100\% & Efficient for large-scale regular structures. \\
Fractal Heat Sink & 100\% & 91\% & 78\% & N/A & Strong capability for innovative, algorithmic design. \\
\bottomrule
\end{tabular}}
\end{table}

Despite these promising results, the MDO Agent is not without limitations. The primary challenge lies in the LLM's current capabilities in complex spatial reasoning and deep domain understanding. While highly effective for standard procedures, the agent still required human-in-the-loop assistance for tasks involving intricate geometric selections or highly nuanced engineering judgments. This indicates that enhancing the agent's autonomy in these specific areas is a critical direction for future work. Additionally, improving the synergy between agents and expanding compatibility with a wider range of engineering software will be key to broader industrial adoption.

\section{Conclusion and Future Work}
\label{sec:conclusion}

This paper introduced a preliminary, semi-automated Multidisciplinary Design and Optimization (MDO) Agent driven by Large Language Models. The agent organizes a closed-loop workflow that links conceptual design generation, parametric Text-to-CAD modeling with visual self-correction, performance verification, and structural optimization. By integrating retrieval-augmented knowledge grounding, natural-language–driven parametric modeling, and ReAct-style tool use for engineering software, the framework seeks to mitigate the limitations of conventional, expert-centric workflows and to shorten iteration cycles under human oversight. Validation on three representative cases—a gas-turbine blade, a machine-tool column, and a fractal heat sink—shows that the agent can complete the pipeline from natural-language intent to verified and optimized designs with reduced manual scripting and setup effort, while encouraging broader exploration of design concepts. We view this as an initial step toward practical, human–AI collaborative mechanical engineering.

Looking ahead, several directions appear particularly promising. First, domain-specialized LLMs aligned to mechanical engineering terminology, units, constraints, and document styles could improve semantic fidelity and the quality of generated solutions. Second, stronger multimodal spatial reasoning—incorporating sketches, point clouds, and legacy CAD—may alleviate the current bottlenecks in geometric understanding, especially for fine-grained boundary-condition definition and entity selection. Third, richer human–agent interaction is needed to support interactive co-design beyond text and code, for example by combining freehand sketching, voice input, and interpretable visual feedback with explicit handoff and rollback mechanisms. Finally, coupling design with manufacturability (DFM) and other downstream lifecycle considerations will be essential to move from proof-of-concept demonstrations toward dependable, vertically customized MDO systems, enabling early manufacturability assessment and optimization under realistic process and cost constraints.

\section{Funding Sources}
The work was supported by the National Key Research and Development Program of China (2022YFE0196400), the National Natural Science Foundation of China (52305288), Zhejiang Provincial Natural Science Foundation of China (Grant no. LXR22E060001), and the Key Research and Development Program of Zhejiang (2023C01166).

\section{Author Contributions}
Bingkun Guo: Methodology, Investigation, Writing original draft; Wentian Li: Methodology, Investigation, Software; Jiaqi Luo: Conceptualization, Investigation, Manuscript review and editing; Zibin Yu: Resources, Data curation; Dalong Dong: Data curation, Validation; Xiaojian Liu: Conceptualization, Project administration; 
Shuyou Zhang: Conceptualization; Funding acquisition; Yiming Zhang: Supervision, Methodology, Writing original draft.

\section*{Declaration of Competing Interest}
The authors declare that they have no known competing financial
interests or personal relationships that could have appeared to influence the work reported in this article.

\section*{Data Availability Statement}
The data that support the findings of this study are available from the authors upon reasonable request.

\bibliography{Refrence}

\begin{thebibliography}{10}
\expandafter\ifx\csname url\endcsname\relax
  \def\url#1{\texttt{#1}}\fi
\expandafter\ifx\csname urlprefix\endcsname\relax\def\urlprefix{URL }\fi
\expandafter\ifx\csname href\endcsname\relax
  \def\href#1#2{#2} \def\path#1{#1}\fi

\bibitem{pereira2022review}
J.~L.~J. Pereira, G.~A. Oliver, M.~B. Francisco, S.~S. Cunha~Jr, G.~F. Gomes, A review of multi-objective optimization: methods and algorithms in mechanical engineering problems, Archives of Computational Methods in Engineering 29~(4) (2022) 2285--2308.

\bibitem{wang2024computer}
T.~Wang, D.~Wu, Computer-aided traditional art design based on artificial intelligence and human-computer interaction, Computer-Aided Design and Applications 21~(1).

\bibitem{zhao2024explainability}
H.~Zhao, H.~Chen, F.~Yang, N.~Liu, H.~Deng, H.~Cai, S.~Wang, D.~Yin, M.~Du, Explainability for large language models: A survey, ACM Transactions on Intelligent Systems and Technology 15~(2) (2024) 1--38.

\bibitem{annepaka2025large}
Y.~Annepaka, P.~Pakray, Large language models: a survey of their development, capabilities, and applications, Knowledge and Information Systems 67~(3) (2025) 2967--3022.

\bibitem{kartashov2025large}
N.~Kartashov, N.~N. Vlassis, A large language model and denoising diffusion framework for targeted design of microstructures with commands in natural language, Computer Methods in Applied Mechanics and Engineering 437 (2025) 117742.

\bibitem{vaithilingam2022expectation}
P.~Vaithilingam, T.~Zhang, E.~L. Glassman, Expectation vs. experience: Evaluating the usability of code generation tools powered by large language models, in: Chi conference on human factors in computing systems extended abstracts, 2022, pp. 1--7.

\bibitem{chang2024survey}
Y.~Chang, X.~Wang, J.~Wang, Y.~Wu, L.~Yang, K.~Zhu, H.~Chen, X.~Yi, C.~Wang, Y.~Wang, et~al., A survey on evaluation of large language models, ACM transactions on intelligent systems and technology 15~(3) (2024) 1--45.

\bibitem{mckinzie2024mm1}
B.~McKinzie, Z.~Gan, J.-P. Fauconnier, S.~Dodge, B.~Zhang, P.~Dufter, D.~Shah, X.~Du, F.~Peng, A.~Belyi, et~al., Mm1: methods, analysis and insights from multimodal llm pre-training, in: European Conference on Computer Vision, Springer, 2024, pp. 304--323.

\bibitem{naveed2023comprehensive}
H.~Naveed, A.~U. Khan, S.~Qiu, M.~Saqib, S.~Anwar, M.~Usman, N.~Akhtar, N.~Barnes, A.~Mian, A comprehensive overview of large language models, ACM Transactions on Intelligent Systems and Technology.

\bibitem{gao2024large}
C.~Gao, X.~Lan, N.~Li, Y.~Yuan, J.~Ding, Z.~Zhou, F.~Xu, Y.~Li, Large language models empowered agent-based modeling and simulation: A survey and perspectives, Humanities and Social Sciences Communications 11~(1) (2024) 1--24.

\bibitem{machado2019parametric}
F.~Machado, N.~Malpica, S.~Borromeo, Parametric cad modeling for open source scientific hardware: Comparing openscad and freecad python scripts, Plos one 14~(12) (2019) e0225795.

\bibitem{cui2024scaled}
Y.~Cui, S.~Ya, C.~Song, A scaled boundary finite element approach for elastoplastic analysis and implementation in abaqus, Computer Methods in Applied Mechanics and Engineering 432 (2024) 117349.

\bibitem{sheng2025isogeometric}
J.~Sheng, X.~Wei, Isogeometric topology optimization of thin-walled structures with complex design domains, Computer Methods in Applied Mechanics and Engineering 444 (2025) 118114.

\bibitem{li2024survey}
X.~Li, S.~Wang, S.~Zeng, Y.~Wu, Y.~Yang, A survey on llm-based multi-agent systems: workflow, infrastructure, and challenges, Vicinagearth 1~(1) (2024) 9.

\bibitem{zhang2024vision}
J.~Zhang, J.~Huang, S.~Jin, S.~Lu, Vision-language models for vision tasks: A survey, IEEE Transactions on Pattern Analysis and Machine Intelligence.

\bibitem{koziolek2024llm}
H.~Koziolek, S.~Gr{\"u}ner, R.~Hark, V.~Ashiwal, S.~Linsbauer, N.~Eskandani, Llm-based and retrieval-augmented control code generation, in: Proceedings of the 1st International Workshop on Large Language Models for Code, 2024, pp. 22--29.

\bibitem{xu2024retrieval}
Z.~Xu, M.~J. Cruz, M.~Guevara, T.~Wang, M.~Deshpande, X.~Wang, Z.~Li, Retrieval-augmented generation with knowledge graphs for customer service question answering, in: Proceedings of the 47th International ACM SIGIR Conference on Research and Development in Information Retrieval, 2024, pp. 2905--2909.

\bibitem{hu2024enhancing}
J.~Hu, W.~Xia, X.~Zhang, C.~Fu, W.~Wu, Z.~Huan, A.~Li, Z.~Tang, J.~Zhou, Enhancing sequential recommendation via llm-based semantic embedding learning, in: Companion Proceedings of the ACM Web Conference 2024, 2024, pp. 103--111.

\bibitem{zhuang2023toolqa}
Y.~Zhuang, Y.~Yu, K.~Wang, H.~Sun, C.~Zhang, Toolqa: A dataset for llm question answering with external tools, Advances in Neural Information Processing Systems 36 (2023) 50117--50143.

\bibitem{wu2024avatar}
S.~Wu, S.~Zhao, Q.~Huang, K.~Huang, M.~Yasunaga, K.~Cao, V.~Ioannidis, K.~Subbian, J.~Leskovec, J.~Y. Zou, Avatar: Optimizing llm agents for tool usage via contrastive reasoning, Advances in Neural Information Processing Systems 37 (2024) 25981--26010.

\bibitem{wang2023recent}
X.~Wang, Y.~Jin, S.~Schmitt, M.~Olhofer, Recent advances in bayesian optimization, ACM Computing Surveys 55~(13s) (2023) 1--36.

\bibitem{yao2023react}
S.~Yao, J.~Zhao, D.~Yu, N.~Du, I.~Shafran, K.~Narasimhan, Y.~Cao, React: Synergizing reasoning and acting in language models, in: International Conference on Learning Representations (ICLR), 2023.

\bibitem{summers2006towards}
J.~D. Summers, A.~Divekar, S.~Anandan, Towards establishing the design exemplar as a cad query language, Computer-Aided Design and Applications 3~(1-4) (2006) 523--534.

\bibitem{shaimi2022ansys}
M.~Shaimi, R.~Khatyr, J.~K. Naciri, Ansys mechanical automation using python for the steady state thermal analysis of fins, in: Proceedings of the World Congress on Mechanical, Chemical, and Material Engineering, Prague, Czech Republic, Vol.~8, Avestia, 2022, pp. HTFF--178.

\bibitem{imambi2021pytorch}
S.~Imambi, K.~B. Prakash, G.~Kanagachidambaresan, Pytorch, Programming with TensorFlow: solution for edge computing applications (2021) 87--104.

\bibitem{farahani2024power}
A.~S. Farahani, H.~Kohandel, H.~Moradtabrizi, S.~Khosravi, E.~Mohammadi, A.~Ramesh, Power generation gas turbine performance enhancement in hot ambient temperature conditions through axial compressor design optimization, Applied Thermal Engineering 236 (2024) 121733.

\bibitem{khater2023effectives}
M.~M. Khater, V.~Govindaraj, Effectives of different shaped dimples on a naca airfoil, Babylonian Journal of Mechanical Engineering 2023 (2023) 29--37.

\bibitem{venugopal2019structural}
P.~R. Venugopal, M.~Kalayarasan, P.~Thyla, P.~Mohanram, M.~Nataraj, S.~Mohanraj, H.~Sonawane, Structural investigation of steel-reinforced epoxy granite machine tool column by finite element analysis, Proceedings of the Institution of Mechanical Engineers, Part L: Journal of Materials: Design and Applications 233~(11) (2019) 2267--2279.

\bibitem{vu2022fractal}
C.~Vu, T.~Truong, J.~Kim, Fractal structures in flexible electronic devices, Materials Today Physics 27 (2022) 100795.

\bibitem{wang2023performance}
H.~Wang, X.~Chen, Performance improvements of microchannel heat sink using koch fractal structure and nanofluids, in: Structures, Vol.~50, Elsevier, 2023, pp. 1222--1231.

\end{thebibliography}

\end{sloppypar}
\end{document}